\def\H        {{$^1$H \/}}

\def\celcius  {{$^\circ$C}}

\newcommand{\ee}[1]{\times10^{#1}}
\newcommand{\mr}[1]{\mathrm{#1}}
\newcommand{\unit}[1]{\,\mathrm{#1}}
\newcommand{\um}{\,\mu{\rm m}}
\newcommand{\us}{\,\mu{\rm s}}
\newcommand{\uT}{\,\mu{\rm T}}

\newcommand{\yH}{\gamma_n}
\newcommand{\ye}{\gamma_e}
\newcommand{\Brms}{B_\mr{rms}}
\newcommand{\xmod}{Y}
\newcommand{\sigf}{\sigma_f}
\newcommand{\Tcp}{T_{2}}

\newcommand{\captionstyle}{\normalfont} 

\documentclass[aps,twocolumn]{revtex4}

\usepackage{graphicx}
\usepackage{bm}
\usepackage{amsmath}
\usepackage{amssymb}
\usepackage{verbatim}
\usepackage[colorlinks=true, pdfstartview=FitV, linkcolor=blue, citecolor=blue, urlcolor=blue]{hyperref} 

\begin{document}

\global\emergencystretch = .1\hsize 

\title{Nanoscale nuclear magnetic resonance with a 1.9-nm-deep nitrogen-vacancy sensor}

\author{M. Loretz$^1$, S. Pezzagna$^2$, J. Meijer$^2$, and C. L. Degen$^1$}
  \email{degenc@ethz.ch} 
  \affiliation{
   $^1$Department of Physics, ETH Zurich, Schafmattstrasse 16, 8093 Zurich, Switzerland.
   $^2$Institute for Experimental Physics II, Department of Nuclear Solid State Physics, Universit\"at Leipzig, Linn\'estr. 5, D-04103 Leipzig, Germany.
	}
\date{\today}

\begin{abstract}
We present nanoscale nuclear magnetic resonance (NMR) measurements performed with nitrogen-vacancy (NV) centers located down to about 2 nm from the diamond surface.  NV centers were created by shallow ion implantation followed by a slow, nanometer-by-nanometer removal of diamond material using oxidative etching in air.  The close proximity of NV centers to the surface yielded large \H NMR signals of up to 3.4 $\mu$T-rms, corresponding to $\sim 330$ statistically polarized or $\sim 10$ fully polarized proton spins in a $(1.8 \unit{nm})^3$ detection volume.  
\end{abstract}


\maketitle

The proposal of diamond magnetometry \cite{degen08,taylor08} and its subsequent demonstration \cite{balasubramanian08,maze08} has received considerable attention for potential applications in nanoscale magnetic resonance imaging and spectroscopy with single nuclear spin resolution \cite{schirhagl14}.  Recently, diamond-based magnetic sensors have enabled detection of \H nuclear magnetic resonance (NMR) from organic molecules deposited on the surface of a diamond chip with a sensitivity of $10^4-10^6$ proton nuclei \cite{mamin13,staudacher13,ohashi13}, which is a roughly one-million-fold improvement compared to conventional NMR \cite{ciobanu02} and on par with magnetic resonance force microscopy \cite{degen09,poggio10}.  Recent advances with diamond sensors were made possible by the controlled positioning of nitrogen-vacancy (NV) centers within less than 20 nm from the diamond surface \cite{ohno12,oforiokai12,ohashi13}.  In order to eventually detect single nuclear spins, NV centers must be moved even closer to the surface in order to pick up the rapidly decaying dipolar field of a single magnetic moment.  
Here, we discuss nanoscale NMR measurements performed with NV centers down to 2 nm from the diamond surface.  These NV centers were created by shallow implantation followed by controlled removal of a few nanometers of diamond material by oxidative etching in air.  The close proximity of NV centers to the surface allowed us to detect as few as $330$ statistically polarized hydrogen nuclei in an organic calibration sample as well as in the adsorbate layer naturally present on the diamond surface.

\begin{figure}[t]
\centering
\includegraphics[width=0.90\columnwidth]{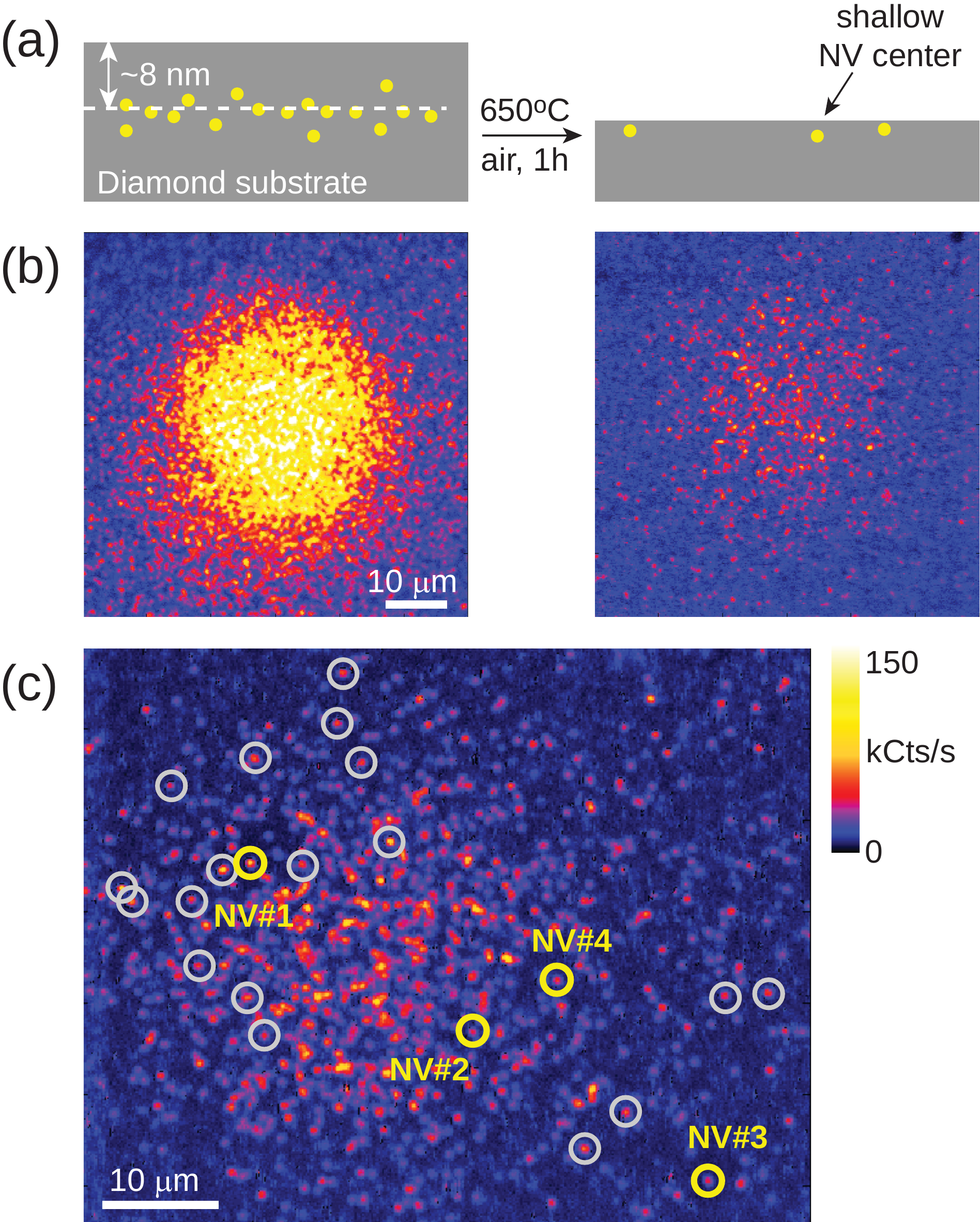}
\caption{\captionstyle
(a) Sketch showing implanted diamond surface (5 keV $^{15}$N$^+$) before and after 1 hour of oxygen etching in air at 650\celcius.  About 10 nm of diamond material are removed, and remaining NV centers are now very close to the surface.
(b) Photoluminescence maps showing the density of NV centers in the implanted area before and after oxygen etching.
(c) Zoom of the implanted area showing fluorescence of individual NV centers.  Circles indicate single NV centers and yellow circles with labels indicate NV centers that were used for \H NMR detection.
Color bar is photon counts per second.
}
\label{fig:etching}
\end{figure}

The diamond chip used in this study was a commercially available single crystal of electronic grade purity and with a (100) surface orientation \cite{e6}.  The two-side polished chip had dimensions of $2\times2\times0.5\unit{mm^3}$.  The as-received diamond was briefly etched by ArCl plasma \cite{lee08} to remove the first few hundred nanometers of material that were possibly compromised by the polishing. 
NV centers were then created by implantation with $^{15}\mr{N}^+$ ions using an energy of 5 keV and a fluence of $10^{11}\unit{cm^2}$ \cite{pezzagna10} and by subsequent annealing at 850$^\circ$C in high vacuum ($p<2\ee{-7}\unit{mbar}$) for two hours.  The peak depth of created NV centers is about $8\unit{nm}$ with a straggling of $\pm 3\unit{nm}$ according to stopping-range-of-ions-in-matter calculations \cite{ziegler10,toyli10,oforiokai12}.  A photoluminescence measurement, shown in Fig. \ref{fig:etching}(b), confirmed that a large number of NV centers ($\sim 5$ NV$^-$ per $\um^2$) was formed by this procedure.

To realize shallower NV centers we have exploited the slow oxidative etching of diamond at $\sim 600$\,\celcius\ in ambient air \cite{osswald06,riedrich12,rugar13}.  This procedure has previously been applied to tune the dimensions of photonic crystal cavities \cite{riedrich12} and has also been considered for depth profiling of shallow NV centers \cite{rugar13}.  We placed the diamond chip in a filament-heated tube furnace \cite{oven} that was open to air at successively higher temperatures until the density of NV centers was substantially reduced.  We found that etching for 1 hour at $650^\circ$ reduced the original NV density to about 20\% of its original value.  We estimate that this corresponds to a removal of about 10 nm of diamond material.  Since the temperature was monitored right at the filament, the actual temperature at the diamond chip's location was probably slighly lower.  A photoluminescence map after etching is shown in Fig. \ref{fig:etching}(b).

We have used \H NMR of a hydrogen-rich calibration sample deposited on the diamond surface to determine the depth of formed NV centers.  For this purpose, we have covered the diamond chip by microscope immersion oil \cite{oil,staudacher13} as a convenient test sample.  The hydrogen content of the oil was measured by mass spectrometry as $\rho = 6\ee{28}$ hydrogen atoms per m$^3$.  The prepared diamond chip was mounted in a custom-built confocal microscope that incorporated a coplanar waveguide for applying fast microwave pulses \cite{oforiokai12,loretz13}.  Single NV centers were localized by confocal imaging and confirmed by optically-detected magnetic resonance (ODMR) measurements \cite{gruber97}.  The microscope was additionally equipped with a moveable permanent magnet to provide vector magnetic fields up to $\sim 0.3\unit{T}$ for NMR experiments.

We measured the statistical polarization \cite{degen07} of \H spins using a Carr-Purcell-type detection sequence (XY8, Ref's. \cite{gullion90,ryan10,delange11,staudacher13}).  We used optical initialization and readout of the NV center \cite{jelezko04} to monitor the transition probability between the $m_s=0$ and $m_s=-1$ electronic spin states after coherent evolution for a fixed time $T$.  During coherent evolution we applied a periodic sequence of microwave $\pi$ pulses to dynamically decouple the NV center from environmental magnetic noise.  NMR signal detection was achieved by adjusting the pulse spacing $\tau$ such that it exactly coincided with half the periodicity of nuclear Larmor precession.  If the Larmor condition is met, that is, if $\tau=1/(2f_0)$, cumulative phase build-up occurs and transitions between the spin states of the NV center are induced.  The measured signal is then proportional to the transition probability $p$.  For proton spins with a gyromagnetic ratio of $\yH = 42.57\unit{MHz/T}$ and in a field of $B_0 = 180\unit{mT}$, the Larmor frequency is about $f_0 = 7.7\unit{MHz}$ and the pulse spacing is about $\tau = 65\unit{ns}$.  Experiments were typically averaged over one million measurements to obtain better statistics.

\begin{figure}[t]
\centering
\includegraphics[width=0.95\columnwidth]{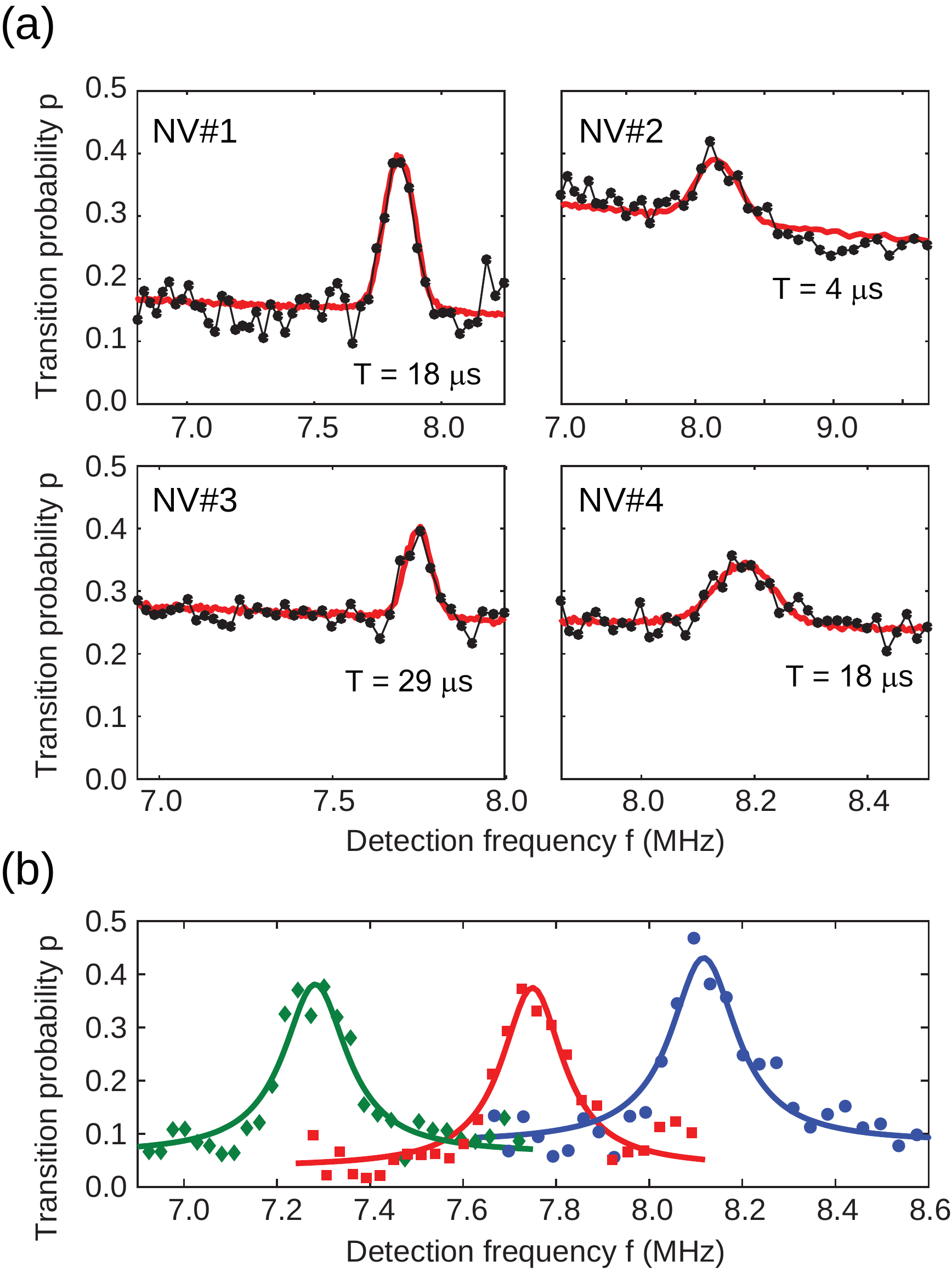}
\caption{\captionstyle
(a) \H NMR of the organic calibration sample recorded using the XY8 protocol. Dots show experimental data and red curves are numerical simulations.
Total evolution time $T$ is given with each spectrum.
Spectra were taken at bias fields between 180-195 mT.  Acquisition times were on the order of a few hours for all spectra.
Further parameters are collected in Table \ref{table:results}.
(b) Spectra recorded by NV\#1 at bias fields $B_0$ of 171 mT (diamonds), 182 mT (squares) and 191 mT (dots) confirm that signals originate from \H nuclear spins.
}
\label{fig:oil}
\end{figure}

Out of about 20 NV centers investigated, we found 4 to show unambiguous proton signals and more defects to show likely signals.  These NV centers are marked in Fig. \ref{fig:etching}(c).  Fig. \ref{fig:oil} shows \H NMR spectra of the organic calibration sample that were recorded by the 4 NV centers with the strongest signals.  We noticed that signals saturated for evolutions times as short as a few microseconds, indicating strong spin noise and a correspondingly small distance of the NV centers to the proton layer on the surface.

We have performed numerical simulations to more precisely estimate the rms-nuclear magnetic field $\Brms$ and the depth $d$ of formed NV centers.  Carr-Purcell-based magnetometry measures the $z$-component of the magnetic field noise produced by the Larmor precession of nuclear spins in the $xy$-plane ($z$ denotes the axis of the NV center and the direction of the external bias field).  The rms-squared nuclear field $\Brms$ for this situation can be analytically calculated by integration over nuclear dipoles (see, e.g., Ref's \cite{rosskopf13,staudacher13}), 
\begin{equation}
\Brms^2 = \frac{5\mu_0^2 h^2\yH^2 \rho}{1536\pi d^3} = (1.14\unit{\uT nm^3})^2 \times \frac{\rho}{d^3} \ ,
\label{eq:brms}
\end{equation}
where $\rho$ is the uniform nuclear spin density, $\mu_0 = 4\pi\ee{-7}\unit{Tm^2/A}$, and $h=6.63\ee{-34}\unit{Js}$.  The transition probability $p$ between the NV center's spin states (the ``signal'') is given by
\begin{equation}
p = \sin^2\left[ \frac{1}{2}\int_0^T dt \ye B_z(t) \xmod(t) \right] \ ,
\label{eq:p}
\end{equation}
where $T=n\tau$ is the total evolution time, $n$ is the number of $\pi$ pulses, $\tau$ is the pulse spacing, and $\ye = 28\unit{GHz/T}$ is the electron gyromagnetic ratio.  $\xmod(t) = (-1)^{[2tf]}$ is the modulation function \cite{delange11} of the multi-pulse detection sequence with detection frequency $f=1/(2\tau)$ and ``[..]'' indicates ``round-to-nearest''.  The random nuclear field $B_z(t)$ is characterized through the magnetic noise spectral density $S(f)$ that is equivalent to the NMR spectrum of the detected nuclei.  We found our signals to be well described by a Gaussian spectral density $S(f) = \Brms^2 (2\pi\sigf^2)^{-1/2}\exp\{-(f-f_0)^2/2\sigf^2\}$, where $\sigf$ is the Gaussian sigma parameter.  For the simulation, we have generated random arrays of $B_z(t)$ and calculated the transition probability $p$ for different detection frequencies $f$.  We have averaged $p$ over many independent samples of $B_z(t)$  and optimized simulation input parameters (namely, $f_0$, $\sigf$, $\Brms$ and $T_2$) by performing a nonlinear regression.  The simulation moreover took finite pulse length (cosine-square-shaped with $\tau_{\pi,\mr{eff}} = 9-13\unit{ns}$) and an exponential $T_2$ decay into account.

Numerical results to experiments and simulations are collected in Table \ref{table:results}.  We found proton spins to produce $\Brms$ between 0.6 and $3.4\unit{\uT}$, which is up to an order of magnitude larger than previous nanoscale NMR experiments \cite{mamin13,staudacher13,ohashi13}.  The depth inferred by Eq. (\ref{eq:brms}) is $d<6\unit{nm}$ for all four NV centers, with the shallowest defect (NV\#2) at $d=1.9\pm 0.2\unit{nm}$.  We believe that this is the shallowest confirmed depth of any NV center reported in the literature.  The simulations also yielded an estimate for the NMR linewidth and the coherence time $\Tcp$ of NV centers.  We noticed that the NMR linewidth of our spectra was large compared to those of typical \H spectra.  We have made this observation before \cite{ohashi13} and attribute it to rapid molecular or spin diffusion through the nanometer detection volume.  We further observed that $T_2$ times (recorded under Carr-Purcell decoupling) are relatively short, on the order of a few to tens of $\us$, which may be due to magnetic surface states \cite{rosskopf13} or due to pulse imperfections.  Spin relaxation measurements \cite{rosskopf13} on NV\#1 showed that $T_1 = 1.4(1)\unit{ms}$ and $T_{1\rho} = 0.43(7)\unit{ms}$ are considerably longer than $\Tcp$.  We finally note that $\Tcp$ increased after removal of the calibration sample.

\begin{figure}[t]
\centering
\includegraphics[width=0.9\columnwidth]{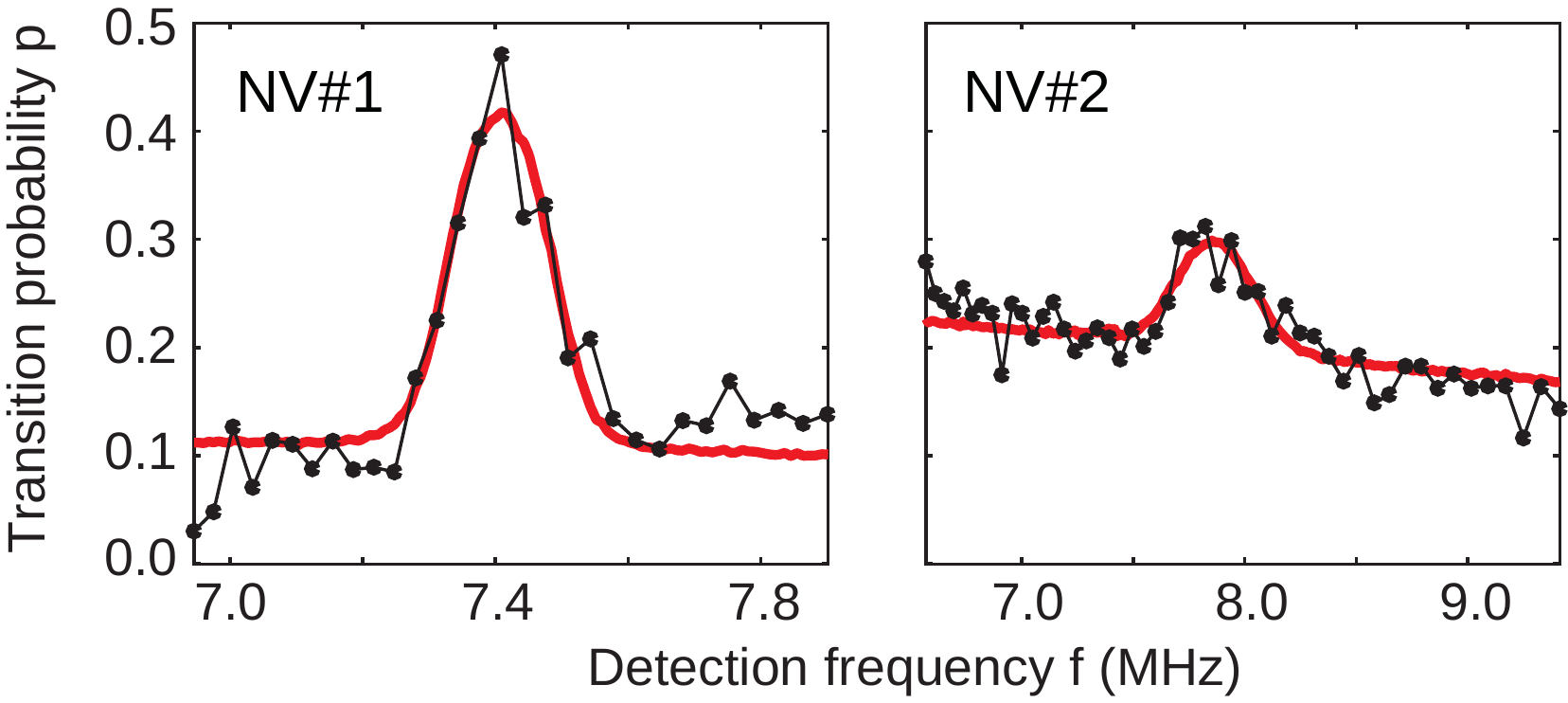}
\caption{\captionstyle
\H NMR of molecular adsorbates naturally present on the diamond surface. Black dots are the data and red lines are simulations.
Experimental parameters are given in Table \ref{table:results}.
}
\label{fig:adsorbates}
\end{figure}

We have found that NV centers show \H NMR signals even in the absence of the organic calibration sample.  Fig. \ref{fig:adsorbates} shows \H NMR spectra measured by NV\#1 and NV\#2 after removal of the sample by, in that order, washing with acetone, washing with methanol, annealing in air at 450\,\celcius\ \cite{osswald06}, and UV-ozone exposure.  Somewhat surprisingly, the NMR signals detected after removing the sample are not much smaller than those recorded from the calibration sample.  We suspect that these signals originate from a thin film of adsorbates on the diamond surface.  The presence of such an adsorption layer is not surprising because it is well known that surfaces that have been exposed to common laboratory atmosphere become covered with a thin film of adsorbed water or hydrocarbons \cite{maier00}.  Moreover, diamond terminating surface groups contain hydrogen \cite{sque06}.  The thickness of the adsorption layer has been measured as $\delta \sim 1\unit{nm}$ by magnetic resonance force microscopy experiments \cite{degen09,mamin09,xue11}.  For our sample we found $\delta\sim0.8\unit{nm}$ for NV\#2 and $\delta>1\unit{nm}$ for NV\#1 by calculating the rms magnetic field as a function of film thickness analogous to Eq. (\ref{eq:brms}), but these $\delta$ carry a large error margin.

The number of proton spins giving rise to the measured signals is quite small.  As an estimate, we have numerically determined the three-dimensional volume above the NV center that generates 70\% of $\Brms$ (or equivalently, 50\% of $\Brms^2$) and counted the number of protons in that volume \cite{mamin13,staudacher13}.  For the thick calibration sample, this volume is approximately $V_{70} \approx (0.98\,d)^3$, where $d$ is the distance of the NV center to the surface.  For NV\#2 that has a depth of $d=1.9\unit{nm}$ the volume is about $V_{70} = (1.8\unit{nm})^3$.  The number of protons in this volume is $N_{70} = \rho V_{70} = 330$.  The number of spins in the thin film sample is smaller as the signal is predominantly produced near the NV sensor.  For NV\#2 and a thickness $\delta=1\unit{nm}$ for the surface film we calculated that $N_{70} = 180$.  Alternatively, we have also compared the measured $\Brms$ to the magnetic dipole field produced by a single proton placed at the optimal location over the NV center.  At a depth of $1.9\unit{nm}$ (NV\#2) the proton dipole field is about $B = 2.06\unit{\uT nm^3}\times d^{-3} = 0.31\unit{\uT}$.  The measured $\Brms$ thus correspond to the magnetic field produced by $N_p = 2.69\uT/0.31\uT \sim 8.6$ fully polarized protons.
\begin{table*}[h]
\centering
\begin{tabular}{cccccccc}
\hline\hline
NV center	& No. of pulses & Total time & Signal			   & Linewidth 	   & Coherence time & Depth 			& No. of \H spins	\\
					& $n$			 			&  $T (\us)$ & $\Brms (\uT)$ & $\sigf$ (kHz) & $T_2 (\us)$		& $d$ (nm)		& $N_{70}$  \\
\hline
\multicolumn{7}{l}{Calibration sample} \\
NV\#1 		&						288 &				  18 & 	 $1.2\pm0.1$ & 	   $45\pm11$ & 	   	 $54\pm4$ & $3.9\pm0.3$ & $3.3\ee{3}$ \\
NV\#2 		&						 64 &  		 		 4 &	 $3.4\pm0.5$ &	 	$124\pm55$ &    $4.5\pm0.2$ & $1.9\pm0.2$ & $330$ \\
NV\#3			&						448 & 				29 & $0.65\pm0.07$ &		 $34\pm 9$ &   		 $41\pm1$ & $5.7\pm0.4$ & $1.0\ee{4}$ \\
NV\#4			&						288 &				  18 & $0.75\pm0.10$ &		 $39\pm14$ & 	  	 $26\pm1$ & $5.2\pm0.5$ & $7.7\ee{3}$ \\
\hline
\multicolumn{7}{l}{Adsorbate layer} \\
NV\#1			&						320 &				  22 &	 $1.3\pm0.2$ &  	 $55\pm14$ &	    $90\pm20$ & ---			    & $^a\,770$ \\
NV\#2			&						 64 &				   4 &	 $2.7\pm0.4$ &    $118\pm56$ &    $8.1\pm0.4$ & ---			    & $^a\,180$ \\
\hline\hline
\end{tabular}
\caption{\captionstyle
Experimental and simulation parameters for the spectra shown in Fig's \ref{fig:oil} and \ref{fig:adsorbates}.
Errors indicate the 95\% confidence interval from the fit.
$\sigf$ is the NMR linewidth given as Gaussian sigma with a corresponding full-width-at-half-height of $2.35\times \sigf$.
$T_2$ is the decoherence time under Carr-Purcell decoupling.
$N_{70}$ is the number of spins that contribute $70\%$ to $\Brms$.
$^a$ assume an adsorbate layer with thickness $\delta=1\unit{nm}$ and proton density $\rho=6\ee{28}\unit{m^{-3}}$.
}
\label{table:results}
\end{table*}

Since the measured NMR signals are strong, the number of spins detected is limited by the spatial resolution of the NV sensor, and not by detection sensitivity.  In order to eventually detect single nuclear spins, the spin density in the sample would have to be diluted, for example by stable isotope labeling or by chemical means.  Diluted nuclei would have the added advantage of narrow NMR resonances that would improve detection sensitivity.  Alternatively, NMR frequencies of adjacent nuclei could be shifted by the application of strong imaging magnetic gradients \cite{degen08,degen09,poggio10,grinolds13}.

In addition to demonstrating detection of small volumes and small numbers of nuclear spins, the simple method to produce very shallow NV centers is the fundamental advance presented here.  Previous nanoscale NMR experiments by diamond magnetometry were done either with isotopically pure substrates \cite{mamin13,ohashi13} or on rather deep defects \cite{staudacher13,mamin13}, with numbers of spins between about $10^4-10^6$.  In contrast, our sample is available commercially and can be prepared easily and with little sophisticated equipment.

This work was supported by the Swiss National Science Foundation through Project Grant $200021\_137520/1$ and through the NCCR QSIT, and by the EU DIADEMS and DARPA Quasar collaborations.
We thank J. Boss, K. Chang, A. Dussaux, F. Jelezko, L. McGuinness, T. Rosskopf, R. Schirhagl, and Y. Tao for experimental support and fruitful discussions.

\end{document}